\documentclass[twoside]{article}
\usepackage{fleqn,espcrc2}
\usepackage{here,graphicx}

\newcommand{\figscale}{0.48}
\newcommand{\hatD}{\hat{D}_{oo}}
\newcommand{\Npoly}{N_{\mathit{poly}}}

\title{An exact Polynomial Hybrid Monte Carlo 
       algorithm for dynamical Kogut-Susskind fermions 
       \thanks{Presented by K-I. Ishikawa}}

\newcommand{\Tsukuba}%
{Institute of Physics, University of Tsukuba, Tsukuba, Ibaraki 305-8571, Japan}

\newcommand{\RCCP}%
{Center for Computational Physics, University of Tsukuba, Tsukuba, Ibaraki 305-8577, Japan}

\newcommand{\ICRR}%
{Institute for Cosmic Ray Research, University of Tokyo, Kashiwa, Chiba 277-8582, Japan}

\newcommand{\KEK}%
{High Energy Accelerator Research Organization (KEK), Tsukuba, Ibaraki 305-0801, Japan}

\newcommand{\YITP}%
{Yukawa Institute for Theoretical Physics, Kyoto University, Kyoto 606-8502, Japan}

\newcommand{\Hiroshima}%
{Department of Physics, Hiroshima University, Higashi-Hiroshima, Hiroshima 739-8526, Japan}
       
\author{
  JLQCD Collaboration: 
  K-I.~Ishikawa\address{\Tsukuba}$^{,}$\address{\RCCP},
  M.~Fukugita\address{\ICRR},
  S.~Hashimoto\address{\KEK},
  N.~Ishizuka$^{\mathrm{a,b}}$,
  Y.~Iwasaki$^{\mathrm{a,b}}$,
  K.~Kanaya$^{\mathrm{a}}$,
  Y.~Kuramashi$^{\mathrm{c}}$,
  M.~Okawa\address{\Hiroshima},
  N.~Tsutsui$^{\mathrm{c}}$,
  A.~Ukawa$^{\mathrm{a,b}}$,
  N.~Yamada$^{\mathrm{c}}$, 
  T.~Yoshi\'{e}$^{\mathrm{a,b}}$ }

\begin{document}

\begin{abstract}
We present a polynomial Hybrid Monte Carlo (PHMC) algorithm as
an exact simulation algorithm with dynamical Kogut-Susskind
fermions.
The algorithm uses a Hermitian polynomial approximation 
for the fractional power of the KS fermion matrix.
The systematic error from the polynomial approximation is removed 
by the Kennedy-Kuti noisy Metropolis test so that the algorithm 
becomes exact at a finite molecular dynamics step size. 
We performed numerical tests with $N_f$$=$$2$ case on several 
lattice sizes. We found that the PHMC algorithm works on
a moderately large lattice of
$16^4$ at $\beta$$=$$5.7$, $m$$=$$0.02$ 
($m_{\mathrm{PS}}/m_{\mathrm{V}}$$\sim$$0.69$)
with a reasonable computational time.

\vspace*{-1.0em}
\end{abstract}

\maketitle


\section{Introduction}
\vspace*{-0.5em}

The low energy QCD dynamics in the real world 
will be understood by lattice QCD with three-flavors of dynamical quarks.
Several efforts have been spent to develop exact numerical algorithms 
with an odd-numbers of the Wilson type quarks~\cite{Nf3}.

The Kogut-Susskind (KS) fermion is an attractive formalism 
since the numerical simulation with much lighter quark 
masses are possible thanks to the remnant chiral symmetry.
Although lattice QCD with the two- or single-flavor KS 
fermions can be defined by taking the fractional power of 
the KS fermion, efficient exact algorithms are not still known.
Approximate algorithms such as 
the $R$-algorithm~\cite{R_Algorithm}
have been used in these cases.


Several exact algorithms are proposed for two-
or single-flavor dynamical KS fermions~\cite{RHMC,Hasenfratz_Knechtli}.
In this paper, we further study 
the idea by Horv\'{a}th \textit{et al.}~\cite{RHMC} 
in the case of the polynomial Hybrid Monte Carlo (PHMC) algorithm.
We develop two types of the PHMC algorithm depending on the choice
of the effective action derived through the polynomial approximation.
We compare the computational cost of these two PHMC algorithms.
We investigate the property of the algorithm on several lattice sizes
in $N_f$$=$$2$ case.
We found that our algorithm shows satisfactory efficiency on 
a $16^4$ lattice with $\beta$$=$$5.7$, $m$$=$$0.02$ 
$(m_{\mathrm{PS}}/m_{\mathrm{V}}$$\sim$$0.69)$.

\vspace*{-0.5em}
\section{Algorithm}
\vspace*{-0.5em}

We construct
two types of the PHMC algorithm, which are refereed to as case A and case B.
Introducing a polynomial approximation and pseudo-fermion field,
the partition function can be generally rewritten in the following form:
\begin{eqnarray}
Z=&&\!\!\!\!\!\!\!\!\!\!\!
    \int {\cal D}U{\cal D}\phi_{o}^{\dag}{\cal D}\phi_{o}
    \ \det[W^{(X)}[\hatD]]^{N_{f}/4}     
    \nonumber\\
 && \ \ \ \ \ \ \ \ \ \ \ \ \ \ \ \ 
    \times e^{-S_{g}[U]-S^{(X)}_{q}[U,\phi_{o}^{\dag},\phi_{o}]},
\label{eq:Effective_Action1}
\end{eqnarray}
where $S_{g}$ is a lattice gauge action, 
$\phi_{o}$ is pseudo-fermion field living only on odd sites.
$S^{(X)}_{q}$ and $W^{(X)}$ are the pseudo-fermion action and 
the correction matrix respectively.
The superscript $(X)$ takes $(A)$ or $(B)$ depending on the type of 
the PHMC algorithm as follows.


\textbf{Case~A:}\
We approximate $\hatD^{-N_f/8}$ by a $\Npoly^{(A)}$ order
polynomial $P_{\Npoly^{(A)}}[\hatD]$. 
The pseudo-fermion action and the correction term become
\begin{eqnarray}
S^{(A)}_{q}[U,\phi_{o}^{\dag},\phi_{o}]&=&
|P_{\Npoly^{(A)}}[\hatD]\phi_{o}|^{2},\nonumber\\
W^{(A)}[\hatD]&=&\hatD(P_{\Npoly^{(A)}}[\hatD])^{8/N_{f}}.
\label{eq:CaseA_Action}
\end{eqnarray}

\textbf{Case~B:}\
We approximate $\hatD^{-N_f/4}$ by an even-order
$\Npoly^{(B)}$ polynomial $P_{\Npoly^{(B)}}[\hatD]$. 
The pseudo-fermion action and the correction term can be written as
\begin{eqnarray}
S^{(B)}_{q}[U,\phi_{o}^{\dag},\phi_{o}]&=&
|Q_{\Npoly^{(B)}}[\hatD]\phi_{o}|^{2},\nonumber\\
W^{(B)}[\hatD]&=&\hatD(P_{\Npoly^{(B)}}[\hatD])^{4/N_{f}},
\label{eq:CaseB_Action}
\end{eqnarray}
where 
$Q_{\Npoly^{(B)}}$ is the $\Npoly^{(B)}/2$ order polynomial 
defined by
$P_{\Npoly^{(B)}}[\hatD]$$=$$|Q_{\Npoly^{(B)}}[\hatD]|^{2}$.

The KS-fermion operator $\hatD$ is even-odd preconditioned as
$\hatD$$=$$\mathbf{1}_{oo}$$-$$\lambda^2 \hat{M}_{oo}$ with
$\lambda^{2}$$=$$2 \Lambda_{\mathrm{max}}$$/$ $(4m^2$$+$$2\Lambda_{\mathrm{max}}^{2})$
and 
$\hat{M}_{oo}$$=$$2 M_{oe}M_{eo}/\Lambda_{\mathrm{max}}^2$$+$$\mathbf{1}_{oo}$.
$\Lambda_{\mathrm{max}}$ is chosen so that the all eigenvalues of $\hat{M}_{oo}$ 
fall into the region $[-1,1]$.
$M_{oe}$ ($M_{eo}$) is the usual KS hopping matrix from even (odd) to 
odd (even) sites.

For both cases, the algorithm takes the following two steps;
(i) perform the HMC algorithm according to the effective action 
Eq.~(\ref{eq:CaseA_Action}) or (\ref{eq:CaseB_Action}),
(ii) when the HMC Metropolis test is accepted, apply
the Kennedy-Kuti noisy Metropolis test to incorporate
the correction term $W^{(X)}[\hatD]]^{N_f/4}$.
Thus we obtain two types of the PHMC algorithm depending on
the choice of the effective action.

The acceptance  probability of the noisy-Metropolis test is defined by
\begin{eqnarray}
&&P_{\mathrm{NMP}}[U\rightarrow U']=\min[1,e^{-dS}],\nonumber\\
&&dS=\zeta_{o}^{\dag}W^{(X)}[\hatD']^{-N_f/4}\zeta_{o}-|\eta_{o}|^2,
\label{eq:dS}
\end{eqnarray}
where $\zeta_{o}$$=$$W^{(X)}[\hatD]^{N_f/8}\eta_{o}$
with a Gaussian noise vector $\eta_{o}$.
$W^{(X)}[\hatD]$ is calculated on an initial configuration and 
$W^{(X)}[\hatD']$ is on a trial configuration
generated by the preceding HMC algorithm.

The fractional power of the correction matrix $W^{(X)}$ 
is taken by the Lanczos based Krylov subspace method
proposed by Bori\c{c}i~\cite{Borici}.
We modified his algorithm suitable to our purpose.
As indicated by Bori\c{c}i we
employ CG based stopping criterion for the Lanczos based method.

\vspace*{-0.5em}
\section{Cost estimate}
\vspace*{-0.5em}

The computational cost is counted as the number of multiplication of 
the hopping matrix to evolve the algorithm unit trajectory.
We employ single leapfrog integration scheme 
for the molecular dynamics (MD) step.
We roughly estimate it as 
\begin{eqnarray}
&&\!\!\!\!\!\!\!\!\!\!\!
   N_{\mathrm{Cost A}}
    =  (2 \Npoly^{(A)}-1)\times N^{(A)}_{\mathrm{MD}} \nonumber\\
&& \ \ \ \ \ + 3\times ((8/N_f)\times \Npoly^{(A)} +1 ) \times N^{(A)}_{\mathrm{CG}},
\nonumber\\
&&\!\!\!\!\!\!\!\!\!\!\!
   N_{\mathrm{Cost B}}
    =  (  \Npoly^{(B)}-1)\times N^{(B)}_{\mathrm{MD}} \nonumber\\
&& \ \ \ \ \ + 3\times ((4/N_f)\times \Npoly^{(B)} +1 ) \times N^{(B)}_{\mathrm{CG}},
\nonumber
\end{eqnarray}
where 
$N_{\mathrm{MD}}$ is the number of MD step and
$N^{(X)}_{\mathrm{CG}}$ the number of iteration of CG algorithm. 
The CG algorithm is used to generate $\phi_{o}$ 
with the global heat-bath method and in the Lanczos based algorithm 
for the noisy Metropolis test.

Now we compare the costs by specifying $\Npoly^{(A)}$ and $\Npoly^{(B)}$.
For this purpose we employ the Chebyshev polynomial approximation;
\begin{equation}
  x^{-s}=(1+\lambda^{2} y)^{-s}\sim 
P_{\Npoly}[x]=\sum_{i=0}^{\Npoly} c_{k}T_{k}[y],
\end{equation}
where $T_{k}$ is the $k$-th order Chebyshev polynomial, $x$$=$$\hatD$ and 
$y$$=$$-\hat{M}_{oo}$ can be read of.
The coefficients $c_{k}$ are calculated as usual. 

\begin{figure}[tb]
\centering
\includegraphics[scale=\figscale,clip]{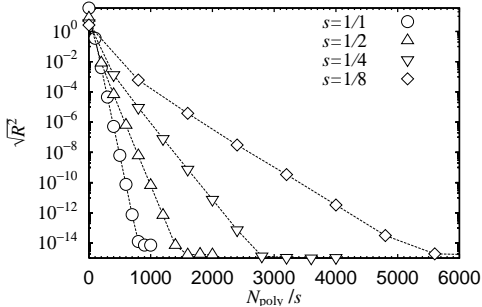}
\vspace*{-3em}
\caption{$\Npoly/s$ dependence of the integrated residual $\sqrt{R^2}$
with $1-\lambda^{2}=1/1000$.}
\vspace*{-2em}
\label{fig:1}
\end{figure}

Figure~\ref{fig:1} shows the cost $\Npoly/s$ dependence
of the integrated residual defined by
$\sqrt{R^2}$$=$$(
   \int_{-1}^{1}dy |  x (P_{\Npoly}[x])^{1/s}-1 |^{2}
  )^{1/2}$
for each $s$. 
We observe that as decreasing $s$ by factor $1/2$ 
the cost $\Npoly/s$ increases by factor $2$ at
a constant $\sqrt{R^{2}}$. 
This is nothing but 
$\Npoly$ does not depend on the choice of $s$ and
we obtain $\Npoly^{(A)}$$\sim$$\Npoly^{(B)}$ at a 
constant approximation level. 

Using this relation and assuming 
$N_{\mathrm{CG}}^{(A)}$$=$$N_{\mathrm{CG}}^{(B)}$ and
$N_{\mathrm{MD}}^{(A)}$$=$$N_{\mathrm{MD}}^{(B)}$, we find
$N_{\mathrm{Cost A}}$$\sim$$ 2 N_{\mathrm{Cost B}}$.
We employ the Chebyshev polynomial and 
the plaquette gauge action, and apply the case B 
PHMC algorithm for numerical simulations.

\vspace*{-0.5em}
\section{Results}
\vspace*{-0.5em}

\begin{figure}[tb]
\centering
\includegraphics[scale=\figscale,clip]{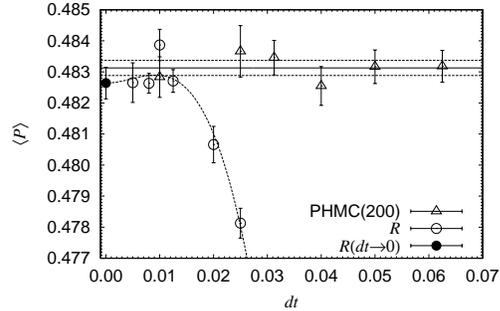}
\vspace*{-3em}
\caption{MD step size $dt$$=$$1/N_{\mathrm{MD}}$ dependence of 
the averaged plaquette 
$\langle P\rangle$ on the small size lattice. 
$\Npoly$$=$$200$, $N_{\mathrm{MD}}$$=$$25$, and 
$\Lambda_{\mathrm{max}}$$=$$2.37$ are employed for the PHMC.}
\vspace*{-2em}
\label{fig:2}
\end{figure}

Figure~\ref{fig:2} shows the MD step size dependence of the averaged 
plaquette on a $8^3\times 4$ lattice at $\beta$$=$$5.26$ and $m$$=$$0.025$.
The results with the PHMC algorithm of $\Npoly$$=$$200$ 
do not depend on $dt$ as it should be and produce the consistent 
result to that in the zero MD step size limit of the $R$-algorithm.
Although we do not show the results on the $\Npoly$ dependence,
the results are independent of the choice of $\Npoly$.

In Table~\ref{tab:1} we show the numerical results on 
a $16^4$ lattice at $\beta$$=$$5.7$ and $m$$=$$0.02$.
The averaged plaquette are independent of $\Npoly$ and consistent 
with each other as expected.
On the other hand, the $R$-algorithm yields 
$\langle P\rangle$$=$$0.577261(49)$~\cite{OKAWA}, which differs from ours 
by $\sim$$2\sigma$. This indicates a potential 
systematic error for the $R$-algorithm at finite MD step size.
The computational time for $\Npoly$$=$$300$ with $m$$=$$0.02$ 
(which corresponds to $m_{\mathrm{PS}}/m_{\mathrm{V}}$$\sim$$0.69$~\cite{OKAWA}) 
was measured as 112 sec. to achieve unit trajectory with 14 GFlops sustained 
speed of SR8000 at KEK.

\begin{table}[t]
  \begin{center}
    \caption{Numerical results on 
a $16^4$ lattice at $\beta$$=$$5.7$ and $m$$=$$0.02$.
    $\Lambda_{\mathrm{max}}$$=$ $2.28$ is employed.
}
    \label{tab:1}
{\scriptsize
  \begin{tabular}{|c|ccc|}\hline
   $N_{\mathrm{poly}}$ &        300 &        400 &        500  \\
$[dt,N_{\mathrm{MD}}]$ &$[0.02,50]$ &$[0.02,50]$ &$[0.02,50]$  \\
   Traj.               &  1700      &   1050     &    800      \\
    $\langle P\rangle$ &0.577099(46)&0.577130(46)&0.577023(43) \\\hline
    \end{tabular}
}
  \end{center}
  \vspace*{-3em}
\end{table}

Consequently we conclude that the PHMC algorithm we constructed
works on a moderately large lattice size with rather heavy quark masses
with reasonable computational cost.
The algorithm also works with a single-flavor fermion.
We emphasize that because the PHMC algorithm is exact one, 
it must be a promising algorithm for future realistic simulations.

\vspace*{3mm}
This work is supported by the Supercomputer Project No.~79 (FY2002) of High
Energy Accelerator Research Organization (KEK), and also in part by the
Grant-in-Aid of the Ministry of Education 
(Nos.~11640294,
12304011,
12740133,
13135204,
13640259,
13640260).
N.Y. is supported by the JSPS Research Fellowship.

\end{document}